# Non-equilibrium spin dynamics in the temperature and magnetic field dependence of magnetization curves of ferrimagnetic $Co_{1.75}Fe_{1.25}O_4$ and its composite with $BaTiO_3$

R.N. Bhowmik[*1], and R. Ranganathan[2]

[1]Department of Physics, Pondicherry University, R. V Nagar, Kalapet, Pondicherry-605014, India.

Condensed Matter Physics Division, Saha Institute of Nuclear Physics, 1/AF Bidhannagar, Kolkata-700064

[*]Corresponding author: Tel.: +91-9944064547; E-mail: rnbhowmik.phy@pondiuni.edu.in

**Abstract**

A comparative study of the non-equilibrium magnetic phenomena (magnetic blocking, memory, exchange bias and aging effect) has been presented for ferrimagnetic $Co_{1.75}Fe_{1.25}O_4$ (CFO) and its composite with non-magnetic $BaTiO_3$ (BTO). Synchrotron X-Ray diffraction patterns have confirmed coexistence of CFO and BTO structures in composite, but magnetic spin dynamics have been remarkably modified. The blocking phenomenon of ferrimagnetic domains below the room temperature has been studied by different modes of (zero field cooled and field cooled) magnetic measurements in collaboration with magnetic fields ON and OFF modes and time dependent magnetization. The applications of unconventional protocols during time dependent magnetization measurement at different stages of the temperature and field dependence of the magnetization curves have been useful to reveal the non-equilibrium dynamics of magnetic spin order. The applying of off-field relaxation experiments has made possible to tune the magnetic state and coercivity of the systems. The role of interfacial coupling between magnetic and non-magnetic particles has been understood on different magnetic phenomena (meta-stable magnetic state, exchange bias and memory effect) by comparing the experimental results of $Co_{1.75}Fe_{1.25}O_4$ spinel oxide and it's composite with $BaTiO_3$ particles.

# 1. Introduction

The non-equilibrium spin dynamics in magnetic materials strongly dependent on spin disorder and manifested by many unusual magnetic phenomena, e.g., spin glass, super-spin glass /cluster spin glass, superparamagnetic blocking, exchange bias, domain wall pinning, memory and training effect [1-7]. Each of these phenomena has their own characteristics. The spin glasses are defined by a typical competition between ferromagnetic (FM) and antiferromagnetic (AFM) exchange interactions and frustration of the spins in lattice structure. The spin dynamics below a characteristic freezing temperature becomes slow due to increasing inter-spin interactions. The superparamagnetic blocking of non-interacting magnetic particles (group of spins) occurs below a typical temperature due to relaxation of the particles along their local anisotropy axes. Taking into account the existence of strong inter-particle interactions, the freezing of nanoparticles assembly is defined as super-spin glass or cluster-spin glass [3,7-8]. It is practically difficult to distinguish the features of super-spin glass from superparamagnetic blocking in magnetic nanoparticles, having finite inter-particle interactions, and distribution in size and anisotropy. In such systems, the aging effect (relaxation phenomenon) plays an important role in determining spin dynamics below their freezing/blocking temperature. The magnetic exchange bias effect was primarily modeled for FM and AFM bi-layers [9], but it has been found in many particulate systems where interfacial exchange coupling between FM (core) and weak FM/AFM (shell) structure control the shape of magnetic hysteresis loop [2, 10-11]. The memory effect is another form of non-equilibrium spin dynamics, where new spin configuration/meta-stable state achieved during intermediate stops of zero field or field cooled magnetization curves can be retrieved during re-heating process [1-3]. The memory effect has been observed in a wide range of materials, irrespective of strongly interacting [12-13] and non-interacting spin systems [14-15].

The artificially designed ferrimagnetic-ferroelectric composite and hetero-structured spin systems also showed exchange bias and memory effect [10-11, 16-18]. The exchange bias effect dominates at lower temperatures and memory effect dominates at higher temperatures [10, 13, 19], and both are not free from spin glass freezing, superparamagnetic blocking, anisotropy and domain wall pinning effect. The training effect, on the other hand, is related to an irreversible change in spin structure pinned at domain walls or at the interfaces of FM-AFM structure or at the interfaces of ferromagnetic and ferroelectric systems [20-21]. The disorder induced by coexisting crystalline phases also played role on spin dependent electronic conductivity [22]. Apart from basic understanding, the study of non-equilibrium spin dynamics is useful for applications of strongly interacting electronic spin systems, such as random alloy [3, 7], perovskite [2, 6, 15], and spinel ferrite [3-5], in spin valves, spins filter, read-writing devices, magneto-resistive random access memories, sensors and magnetic switches [23-24]. This requires an effective strategy for tuning the ferro/ferrimagnetic parameters by controlling the effects of spin disorder inside the domains or at interfaces of the composite materials.

The present work focuses on spinel ferrites, which are defined by a general formula unit $AB_2O_4$, where cations occupy the tetrahedral (A) and octahedral (B) coordinated lattice sites with anions ($O^{2-}$) at fcc positions of the lattice structure. In long range ferrimagnetic (FiM) spinel ferrite, antiferromagnetic (AFM) superexchange interactions between A and B site moments (J(A-O-B)) are expected to be strong in comparison to intra-sublattice interactions (J(B-O-B)) and (J(A-O-A)) [25]. In this work, we will study the effects of intrinsic disorder in ferrimagnetic $Co_{1.75}Fe_{1.25}O_4$ particles [26] and extrinsic spin disorder (interfacial effect) in its composite with non-magnetic $BaTiO_3$ [27] to control the non-equilibrium magnetic phenomena, e.g., exchange bias, memory and aging effect.

## 2. Experimental

### 2.1. Material Preparation

The material preparation and characterization of the $Co_{1.75}Fe_{1.25}O_4$ (CFO) ferrite and its composite with $BaTiO_3$ (BTO) were described in earlier works [26-27]. The ferrite powder was prepared by chemical co-precipitation route and thermal annealing at 800 $^0$C (CF80) and 900 $^0$C (CF90) for 2 hrs. The CF80 sample formed bi-phased cubic spinel structure, unlike single phase structure in CF90 sample. The composite sample CF80_BTO was prepared by mixing of CF80 ferrite and BTO powders with mass ratio 50:50, and final heat treatment was performed at 1000 $^0$C for 4 hrs. Synchrotron X-ray diffraction pattern confirmed the coexistence of cubic spinel structure of CFO and tetragonal phase of BTO in the composite CF80_BTO sample without any intermediate phase formation. Interestingly, bi-phased nature of CF80 sample (as seen from split of X-ray diffraction peaks of cubic spinel phase) disappeared in CF80_BTO composite. The spin structure in CF90 ferrite and CF80_BTO composite samples are schematically modeled in Fig. 1(a-b) and origin of the spin disorder for non-equilibrium spin dynamics are summarized below. The single phase ferrite sample CF90 is modeled as consisting of average particle size ~ 40 nm and each magnetic particle is assumed to be consisting of core-shell spin structure [10, 19]. The core (interior) part is consisting of more than one domain (multi-domain structure). The spins inside each domain are ferrimagnetically (↑↓↑↓) ordered and disordered or pinned at the domain-walls. Effectively, the shell (outer) part of a particle spreads over few lattice parameters whose length is more than domain-wall thickness and spins therein are more disordered than the core spins. The magnetic exchange interactions inside the core are stronger than the shell and particles are strongly interacted in CF90. In case of CFO_BTO composite, the ferrite particle of size ~ 90 nm are dispersed in matrix of BTO of micron sized particles. The presence non-magnetic (NM) BTO particle dilutes the magnetic exchange interactions between two CFO (FiM) particles and it increased ferrimagnetic softness in composite sample [27]. The interfacial exchange interactions are affected by possible magneto-electric coupling [21, 28] and hidden exchange coupling [29] between FiM and ferroelectric (FE) BTO particles. Although, both CF90 and CFO_BTO are hetero-structured spin systems, but the nature of interfacial spin disorder is different. In case of hetero-structured spin systems, the time evolution of spin vector inside an ordered magnetic domain can be re-written as $\frac{d\vec{M}}{dt} = -\gamma_0 (\vec{M} \times \vec{H}_{ext}) + \frac{\alpha}{M_S} (\vec{M} \times \frac{\partial \vec{M}_{eff}}{\partial t})$, where $\frac{\partial \vec{M}_{eff}}{\partial t} = \frac{M_S}{\alpha} \vec{H}_{int} + \frac{\partial \vec{M}}{\partial t}$ [30]. A competition between the free spin torque under external field (first term) and intrinsic damping torque (second term) under internal field and meta-stable states determines the relaxation/orientation of spin vectors towards its nearest new magnetic state. The internal field is controlled by spin disorder, frustration and inter-particle interactions. In case of CF90 sample, the spin disorder is contributed by intrinsic disorder at core (ordered FiM) and extrinsic disorder at shell (disordered FiM). In the spinel ferrite $Co_{1.75}Fe_{1.25}O_4$, intrinsic spin disorder is expected due to distribution of magnetic moment and magneto-crystalline anisotropy of the cations among A and B sites of the spinel structure ($Fe^{3+}$ ions at A and B sites in high spin state and low anisotropic, $Co^{2+}$ ions at A and B sites in high spin state and highly anisotropic, and $Co^{3+}$ ions at B sites are non-magnetic and isotropic) [25]. In case of CFO_BTO composite, additional extrinsic spin disorder is introduced at the interfaces of shell (disordered FiM of CFO) and shell (non-magnetic and ferroelectric-BTO). This is produced due to structural and magnetic mismatch at the interfaces of two phases [31]. Hence, the change of both external magnetic field (ON/OFF) and internal field control the time response of magnetization in the temperature and field dependence of magnetization curves. The basic difference is that the magnetization will be well below of the saturation level in case of the temperature dependence of the magnetization curves, where as the magnetization will be close to the saturation level (high magnetic state) in case of the field dependence of magnetization curves at the starting of relaxation process.

**2.2. Measurement protocols**

Physical property measurement system (PPMS-EC2, Quantum Design, USA) was used for magnetic measurements. The temperature dependence of magnetization was recorded using zero field cooled (ZFC) and field cooled (FC) modes with conventional and unconventional protocols (PCs). The PC1 is a conventional ZFC mode (Fig. 1(c)), where the sample was cooled from 330 K to 10 K in the absence of external magnetic field or applying a small field to maintain the residual magnetization close to zero during cooling. This was followed by magnetic measurement at set (constant) magnetic field while temperature of the sample is warming up to 300 K/330 K. The PC2 is the conventional FC mode (Fig. 1(d)), where the sample was cooled from 300 K/330 K to 10 K in the presence of constant magnetic field. The magnetization was recorded during field cooling (MFCC(T)) of the sample from higher temperature or warming (MFCW(T)) of the sample from 10 K to 300 K/330 K without changing the field that was applied during pre-cooling down to 10 K. The conventional (ZFC and FC) measurement protocols provide general features (magnetic blocking and anisotropy effect) of the magnetic particles. The non-equilibrium spin dynamics (memory and aging effect) were examined by adopting few unconventional protocols to record time dependent magnetization during intermediate stop on temperature and field dependence of magnetization [M(T, H, $t_w$)] curves [2, 5, 14-15]. We followed FC protocol (PC3) (Fig. 1(e)) for studying the memory effects. In FC-PC3, MFCC(T) curve was recorded with intermediate stops at 250 K, 150 K and 50 K by switching off the cooling field for time $t_w$. The M($t_w$) data at the stopping temperature were recorded before switching the cooling field again ON and resuming the MFCC(T) measurement on lowering the temperature down to 10 K. After reaching the temperature 10 K, the MFCW(T) curve was recorded from 10 K to 300 K without changing the cooling field and without intermediate stops. The PC4 is the conventional field dependence of magnetization (M(H)) measurement (Fig. 1(f)), pre-cooled under ZFC and FC modes from 300 K to the set temperature. The shift of FC-M(H) loop with respect to ZFC-M(H) loop can be used to study exchange bias effect. The protocol PC5 in Fig. 1(g) is the superposition of PC4 with $M(t_w)$ measurement, where the M (T = constant, H, $t_w$) curve was recorded by varying the magnetic field with an intermediate stop for waiting time ($t_w$) at magnetic field to zero or before coercive field point in negative field axis and $M(t_w)$ data were recorded. After $M(t_w)$ measurement, the M(H) measurement is continued in negative field side. The steps of M(H) measurements are repeated with different $t_w$ values. The protocol PC6 in Fig. 1(h) is similar to the protocol PC5. The only exception is that $M(t_w)$ measurements were carried out at multiple points (at zero field or points close to coercive fields both on negative and positive field axes) of the M(H) curves.

## 3. Results and discussion

### 3.1. Temperature and field dependent magnetization [M(T,H, $t_w$)] for CF90 sample

First, we show basic properties of the temperature dependence of magnetization in CF90 sample. The MZFC(T) and MFCW(T) curves at +500 Oe (Fig. 2(a)) were measured using PC1 and PC2. The MZFC(T) curve exhibits magnetic blocking temperature ($T_m$) at ≥ 300 K. A wide bifurcation between MZFC(T) and MFCW(T) curves below $T_m$, where MZFC curve decreased rapidly below $T_m$ and becomes nearly temperature independence below 150 K, and MFC curve slowly increased. The behavior shows high anisotropic effect at low temperatures. To overcome the anisotropic effect, we measured MZFC(T) curves by increasing the magnetic fields up to ± 2 kOe (Fig. 2(b)). The MZFC(T) curves showed more or less symmetric response of the spin-clusters under field reversal. Magnitude of the MZFC(T) curves increased with a shift of peak position to low temperature on increasing the magnetic field. In highly anisotropic sample, the energy density in ferrimagnetic state in the presence of magnetic field is $E = K_H + K_A$, where $K_H$ ($= \pm M_{sat} H \cos(\theta - \varphi)$) is the Zeeman energy and $K_A$ ($= K_{eff}(T) \sin^2\theta$) is the crystalline anisotropy energy [20]. The MZFC(T) curves below 150 K are also not significantly affected within ± 2 kOe, except some minor differences. It indicates that Zeeman energy is not enough in this field range to overcome the anisotropy energy and domain-wall pinning effect controls the shape of magnetization curves [32]. On the other hand, broad peak in MZFC(T) curves describes a distribution of anisotropy in the system and it can be quantified from first order derivative of the MZFC(T) curves (Fig. 2(c)). The peak profile in the dM/dT vs. T curves (Fig. 2(d-e)) was fitted with Lorenzian shape to determine the peak parameters. The intercept of the dM/dT curve on temperature axis (> $T_p$) defines the blocking temperature ($T_m$). The peak temperature ($T_p$) of dM/dT vs. T curve corresponds to the inflection point of the MZFC(T) curve below $T_m$. Fig. 2(f) shows a symmetrically decrease of the peak parameters ($T_p$, width, $T_m$) about the zero point of magnetic field axis with the increase of field magnitude. The increase of peak height, along with decrease of peak width, arises due to field induced clustering of small particles [13]. The $T_m(H)$ curve is fitted with a power law: $T_m(H) = a-bH^n$ (*a* and *b* constants) with exponent *n* ~ 0.25 and ~ 0.21 for positive and negative fields, respectively. The exponent values for $T_p(H)$ curve are ~ 0.29 and ~ 0.27 for positive and negative fields, respectively. The values of *n* in CF90 sample are suggestive of magnetic spin-clusters coexisting in ferrimagnetic state [33]. The M(T) curves show bulk response of a ferrimagnet without much information of local spin dynamics.

In order to get information of local spin dynamics, the memory effect was tested using protocol PC3. Fig. 3(a) shows the corresponding MFCC(T) curve at cooling field +200 Oe with intermediate stops and subsequent MFCW(T) curve. The appearance of kinks in the MFCW(T) curve at the previously intermittent stops (field off condition at 250 K, 150 K and 50 K during MFCC(T) process) suggests a recovery/memory of the magnetic spin states that were imprinted through redistribution of energy barriers during the cooling process. The memory is reduced on lowering the stopping temperatures and negligible at 10 K. The magnetization that is recovered on re-applying the cooling field depends on the response of spins in relaxed/quasi-relaxed state. In a strongly interacted spin-system, an increasing slow down of the spin dynamics on decreasing the sample temperature below its spin freezing/blocking temperature hinder the recovery of initial magnetic state. It leads to a large step in MFCC(T) curve immediately after switching OFF and re-applying (ON) of the cooling field at the temperatures (e.g., 250 K in our case). The step in MFCC(T) decreases as sample temperature decreases far away from its spin freezing temperature. It is due to increasing inter-spins interactions in a strong spin-pinning state (e.g., 50 K). Interestingly, the MFCW(T) curve overshoots the MFCC(T) curve at temperatures above 300 K. It shows in-field growth of magnetization due to non-equilibrium spin state of the magnetic particles below their true blocking temperature (above 300 K). The measurement of MFCC(T) and MFCW(T) at 200 Oe without field-off at intermediate temperatures formed a thermal hysteresis loop (Fig. 3(b)). It is a characteristic feature of first order magnetic phase transition (short range spin order coexists in long range spin order) in the sample [34]. In our sample, the in-field MFCW(T) curve starts with thermal activated de-pinning of the spins that were in pinning in intrinsically disordered ferrimagnetic state at 10 K after completing the MFCC(T) measurement. However, a difference between MFCW(T) and MFCC(T) curves (right

Y axis of Fig. 3(b)) showed a maximum at about 210 K and it marked different spin dynamics at lower and higher temperatures. Magnitude of the difference decreases at higher temperatures due to approaching of spin system towards blocking temperature (less interacting/pinning effect) and at low temperatures due to approaching towards a strongly spin-pinning/interacting regime. On increasing the magnitude of cooling field to 500 Oe (Fig. 3(c)), the memory effect is observed only at 250 K and suppressed at low temperatures (50 K and 150 K). This is due to clustering of smaller magnetic particles and de-pinning of the spins (domain wall motion) at higher magnetic field. In this process, the distribution of exchange interactions and anisotropy barriers related to cluster size distribution is narrowed down. It results in strongly spin-interacting clusters during field cooling process and reduces the memory effect at 500 Oe. On the other hand, spin system switches its magnetic state from high to low immediately after switching off the cooling field. The spins in low magnetic state (non-zero remanent magnetization) relax for sample temperature in blocking state ($T < T_B$) [5-6, 15]. The time dependence of FC-remanent magnetization (M (t, H= 0)) curves have been analyzed by various equations, e.g., stretched exponential form [7], a complicated form of equation that consists of essentially two power law terms [3]. In our sample, M(t) (normalized by initial value $M(t_0)$) curves during field-off condition ($t = t_w = 1500$ s) (Fig. 3(d-e)) are best fitted with a function, consisting of a constant and two exponential decay terms.

$$M(t) = \alpha_0 \pm \alpha_1 \exp(-t/\tau_1) \pm \alpha_2 \exp(-t/\tau_2) \qquad (1)$$

Sign of the pre-factors $\alpha_1$ and $\alpha_2$ is taken as positive and negative to represent the magnetization decay and growth, respectively. Out of the two exponential terms, one represents fast relaxation (initial process) and other one represents a slow relaxation (secondary process at higher times). Similar spin relaxation processes were found in magnetic systems with heterogeneous spin structure [35]. The fit of M(t) data at 50 K with a logarithmic decay $M(t) = \alpha_0 - m*\ln t$ (with $m =$ 0.0002 and 0.0160 at cooling fields 200 Oe and 500 Oe, respectively) represents an extremely slow spin systems and generally represents a distribution of activation energy in spin glass state [1,3, 36]. A comparative fit of the M(t) data during OFF condition of 500 Oe at 250 K (Fig. 3(f)) suggests that logarithmic decay is satisfied for limited portion of the M(t) curves, but equation (1) widely matched to the M(t) curves. Hence, equation (1) is more acceptable in fitting the M(t) curves during field-off condition of M(T) and M(H) measurements.

The non-equilibrium spin dynamics during ZFC-M(H) loop measurement within field ± 70 kOe at 10 K (Fig. 4(a)) can be studied using PC4 and PC5. The M(H) loop under zero field cooled mode was recorded at 10 K using PC4. Next, M(H) measurement between +70 kOe to -10 kOe was repeated 6 times with intermediate wait at 0 Oe to record the $M(t_w)$ curves for different $t_w$. In principle, spins in ferrimagnetic state is expected to relax during waiting, irrespective of sweeping field ON or OFF conditions, if finite amount of disorder coexists in spin order, and it could produce new meta-stable state in the M(H) path. As shown in Fig. 4(b), the M(H) curves between 0 Oe and -10 kOe are extremely sensitive to spin relaxation during $t_w$ at 0 Oe. The M(H) curves after waiting at 0 Oe move upward with the increase of $t_w$ with reference to the first curve (default $t_w$ = 10 s). The $M(t_w)$ curves at 0 Oe (Fig. 4(c)) slowed down for higher $t_w$ and followed equation (1). Fig. 4(d-f) shows the waiting time dependence of the fit parameters ($H_C$, $M_0$, $\alpha_1$, $\tau_1$, $\alpha_2$, $\tau_2$) from M(H) curves (0 Oe to – 10 kOe) and $M(t_w)$ curves at 0 Oe. Coercivity ($H_C$) of the CF90 sample significantly increased (6628 Oe to 6954 Oe) with the increase of $t_w$ from 100 s to 7200 s at 0 Oe, unlike a decrease of the fit parameter $M_0$ (35.3473 emu/g to 35.304 emu/g). This is associated with faster relaxation of initial process (increasing $\alpha_1$ and smaller $\tau_1$) and slower relaxation of secondary process (decreasing $\alpha_2$ and larger $\tau_2$) with the increase of $t_w$. A wide difference between $\tau_1$ and $\tau_2$ confirms the existence of two relaxation mechanisms in the sample.

In order to study the non-equilibrium spin dynamics in FC-M(H) loops, we have recorded M(H) loops at 10 K using FC-PC4 at cooling fields +70 kOe and -70 kOe. The M(H) curve started from field sweeping +70 kOe to -70 kOe and back to +70 kOe for the FC loop (cooling @ +70 kOe) and in reverse way for the FC loop (cooling @ -70 kOe). As shown in Fig. 5(a), the FC loops exhibit widening and shifting along field and magnetization directions with improved squareness in comparison to ZFC loop at 10 K. It occurs due to exchange coupling of hetero-structured spins at the interfaces or frozen in the system that favor ordering along cooling field direction and irreversible under reversal of the field direction [Khur]. The centers ($H_{C0}$, $M_{R0}$) and coercivity ($H_C$) of the FC and ZFC loops have been used to calculate the shift of coercivity ($\Delta H_C = |H_C^{FC} - H_C^{ZFC}|$), exchange bias field ($H_{EB} = H_{C0}^{FC} - H_{C0}^{ZFC}$) and magnetization ($\Delta M_R = M_R^{FC} - M_R^{ZFC}$). The FC loop (@ +70 kOe) is nearly symmetric with minor exchange bias shift ($H_{EB} \sim +8$ Oe) and its $H_C \sim 8295$ Oe. However, a large positive shift of magnetization ($\Delta M_R \sim +1.55$ emu/g) and coercivity ($\Delta H_C \sim +1595$ Oe) are noted with respect to ZFC loop with $H_C \sim 6760$ Oe. As compared in the inset of Fig. 5(a), FC loop (@ +70 kOe) and FC loop (@ -70 kOe) showed similar features, but FC loop (@ -70 kOe) shows higher widening ($H_C \sim 8495$ Oe, $\Delta H_C \sim +1735$ Oe, $\Delta M_R \sim +1.99$ emu/g) and squareness. This means spin dynamics is anisotropic to the reversal of high field cooling and it could be related to spin pinning in ferrimagnetic domains [11]. In order to study the aging effect at intermediate point of the FC-M(H) curves, we repeated M(H) measurement within field range +70 kOe to -10 kOe for 6 times with waiting at -2.5 kOe by adopting PC5. Fig. 5(b) demonstrates that the shape (upward increase) of the M(H) curve in the field range -2.5 kOe to -10 kOe is controlled by spin relaxation at -2.5 kOe during $t_w$ (140 s to 7200 s). As shown in Fig. 5(c), the $M(t_w)$ curves at -2.5 kOe also followed equation (1). The increase of $t_w$ at -2.5 kOe of the FC-M(H) loop (@+70 kOe) has increased the $H_C$ (Fig. 5(d)-left

Y axis), the overall magnetization after -2.5 kOe and fit parameter $M_0$ (Fig. 5(d)-right Y axis). The fit parameters (Fig. 5(e-f)) for fast relaxation and slow relaxation processes showed similar features as observed with $t_w$ in case of ZFC-M(H) loop experiment (Fig. 4(e-f)).

Next, we tested the spin relaxation on the M(H) curves at 150 K, a temperature just above the magnetization blocking temperature ~ 125 K in MZFC(T) curve for field 50 kOe (Fig. 6(a)). At this temperature, domain wall pinning is less effective but magnetic clusters are not free from mutual interactions. The ZFC-M(H) curve was measured by sweeping field from +70 kOe to -6 kOe and intermediate waiting at -1 kOe. After measurement of the first M(H) curve, the field was made to zero and back to +70 kOe before starting the next curve and repeated it 7 times. Fig. 6(b) shows all the relaxation regime of M(H) curves at -1 kOe during waiting time and subsequent field dependent regime (- 1 kOe to -5 kOe). It is noted (Fig. 6(c)) that magnitude of the M(H) curves for H < -1 kOe is systematically suppressed on increasing the waiting at -1 kOe. This trend is in contrast to the increasing increment for similar experiment at 10 K (Fig. 4(b)). The $M(t_w)$ curves in the relaxation regime (inset of Fig. 6(c)) at 150 K also follow equation (1) and fit parameters are shown in Fig. 6(d-f). The $H_C$ has shown a small increment, whereas $M_0$ decreases with the increase of $t_w$. The values of the pre-factors ($\alpha_1$, $\alpha_2$) and time constants ($\tau_1$, $\tau_2$) at 150 K are relatively larger than the values at 10 K. It indicates a faster decay of magnetization at 150 K, where spin dynamics is still slow due to strong intra-cluster spin interactions.

**3.2 Temperature and field dependent magnetization [M (T, H, $t_w$)] for CF80_BTO sample**

Fig. 7(a) shows the features of the MZFC(T) and MFC(T) curves at 500 Oe in composite sample. It is seen that basic magnetic features of the ferrite particles, e.g., blocking temperature ($T_m$) at about 300 K, wide magnetic bifurcation at low temperatures, and a weak temperature dependent MZFC(T) curve below 150 K, are retained in the BTO matrix [27]. MZFC(T) curves (Fig. 7(b)) also showed field induced magnetic changes, including increasing magnetization and shift of the broad maximum to lower temperatures. First order derivative of the MZFC(T) curves (∂MZFC/∂T) at different magnetic fields (Fig. 7(c)) showed an asymmetric shape about the peak temperature ($T_p$), which is the inflection point below the broad maximum of MZFC(T) curves. The peak profile of ∂MZFC/∂T curves were fitted with Lorentzian curve and the peak parameters are shown in Fig. 7(d). The peak temperature ($T_p$) decreases at higher field by following a power law: $T_p(H) = a-bH^n$ with exponent ($n$) ~ 0.31, which is close to that obtained for CF90 sample. It suggests the retaining of the glassy behavior of spin-clusters in composite system [33]. It is noted that peak height of the ∂MZFC/∂T curves initially increased for field up to 2.5 kOe, followed a gradual decrease at higher fields. This corresponds to a minimum peak width at 2.5 kOe, along with an increase of peak width both at lower and higher magnetic fields. The features are consistent to field induced nucleation of small particles by de-pinning the spins at domain wall or at the interfaces of ferrimagnetic and ferroelectric particles (via grain boundary) at low field regime. The increase of broadness in the first order derivative curves for fields higher than 2.5 kOe is attributed to an increase of intrinsic disorder, arising from a competition of anisotropy constants and exchange interactions inside the clusters, where as the reduced peak height is attributed to quasi-saturation states of magnetization curves at higher fields.

The retaining of memory effect of the ferrite particles in composite sample is confirmed from MFCC(T) and MFCW(T) curves (Fig. 8(a-d)), measured at cooling field range 200 Oe to 10 kOe. The MFCW(T) curves showed kinks at the temperatures (250 K, 150 K, 50 K) where field was switched off during FCC mode. The kinks are more pronounced than the CF90 sample. The memory effect in CF80_BTO sample also reduced at low temperatures and at higher magnetic fields, similar to the features in CF90 sample. In Fig. 8 (e-f), we have compared the % drop and relaxation of remanent magnetization for both the samples during field off condition of the MFCC(T) curves. The % of drop represents fraction of the reversible spins in the system immediately after switching off the cooling field. It is 100 % for non-interacting paramagnetic spins and less than 100 % for existence of finite interactions among the spins or cluster of spins. In case of interacting spins, the relaxation component is non-zero and it represents the fraction of irreversible spins in the system that shows aging effect. The % drop is seen to be higher than the relaxation part during waiting time, as schematized in the inset of Fig. 8(e). It is seen that % of drop and relaxation both are drastically reduce on lowering the measurement temperatures from 250 K to 50 K. However, distinct differences can be noted in the off-field properties between CF90 and CF80_BTO composite samples. For example, the % of drop decreased on increasing the cooling field from 200 Oe to 500 Oe in case of CF90 sample, whereas a monotonic increase noted with the increase of cooling fields from 200 Oe to 10 kOe in composite sample. The higher % drop of magnetization indicates weakening of magnetic spin interactions at the interfaces of CFO (ferrimagnetic) and BTO (non-magnetic) particles. At the same time, increasing drop at higher cooling fields is attributed to fast de-nucleation of larger clusters into smaller clusters (composed of magnetic ferrite and non-magnetic BTO particles) on switching off the cooling field. The de-nucleation of the large clusters is slow in CF90 sample due to strong inter-particle interactions. It gives rise to relatively low values of % drop and relaxation at all the measurement temperatures for CF90 sample. Based on the data for cooling field 500 Oe, it can be mentioned that composite sample is consisting of approximately 29% paramagnetic/non–interacting spins and 3% of the interacting spins relaxed during waiting time 1600 s at 250 K. It is reduced to 1 % (paramagnetic spins) and 0.14 % (relaxation of interacting spins) for temperature 50 K. In case of CF90 sample, the paramagnetic spins (22%) and relaxation of the interacting spins (0.44 %) at 250 K are reduced to 0.52 % (paramagnetic spins) and 0.03 % (relaxation of interacting spins) at 50 K. The M($t_w$) curves during cooling field off condition of MFCC(T) measurements were fitted using equation 1 (Fig. 9 (a-d)) and the fit parameters are shown in Fig. 9(e-i). Variation of the parameter M(0) is consistent to the temperature and field dependence of magnetization curves. The fit parameters associated with relaxation processes ($\tau_1$, $\tau_2$, $\alpha_1$ and $\alpha_2$) are less sensitive to higher magnetic fields (5 kOe and 10 kOe) and suggest quasi-equilibrium state due to nucleation of clusters. However, the time constants ($\tau_1$, $\tau_2$) are relatively high at 50 K and further increased for higher cooling fields. It indicates slowing down of the spin dynamics at low temperature due to increasing interactions among the spins/cluster of spins and intra-cluster interactions (domain wall motion) dominate at higher cooling fields. Most importantly, magnitude of the time constants in composite sample is nearly one order less than the values in CF90 sample. This is an evidence of faster relaxation in composite sample due to less inter-cluster exchange interactions.

The variation of coercivity in the composite sample as the function of in-field waiting time on FC-M(H) curve has been examined by using PC5 at 10 K (Fig. 10 (a)) and 150 K (Fig. 10(b)). The sample was first zero field cooled from 300 K to 10 K/150 K and M(H) curve (N = 1) was measured during sweeping of the field from +70 kOe to -20 kOe (10 K) or complete loop was measured at 150 K. After first round measurement, the applied field was made to zero before increasing the temperature to 300 K. Next, the sample was cooled under +70 kOe down to 10 K/ 150 K. After temperature stabilization, the M(H) curve (N = 2) was recorded from +70 kOe to - 20 kOe with an intermediate waiting at - 7 kOe for 10 K and at -2150 Oe for 150 K. The M($t_w$) curve was recorded for different in-field waiting times by repeating the FC-M(H) curve in the field range +70 kOe to -20 kOe. The (normalized) in-field M($t_w$) curves are shown for 10 K (Fig. 10(c)) and for 150 K (Fig. 10(d)). The inset of Fig. 10(a) shows that $H_C$ at 10 K is enhanced in FC-M(H) curve and the $H_C$ is further enhanced by repeating the FC loop with higher waiting time at - 7 kOe. The $t_w$ at 10 K was made high enough to observe an appreciable relaxation. The in-field M($t_w$) curves at 10 K were fitted with logarithmic decay law M($t_w$) = $\alpha_0$ −$m$*ln$t_w$. The inset of Fig. 10(c) shows the decrease of both $\alpha_0$ and slope ($m$) with the increase of waiting time at - 7 kOe of the FC-M(H) curves. In contrast, in-field M($t_w$) curves at 150 K were fitted with exponential law (1). The fitted parameters are not shown in the graph. On the other hand, in-field (-2150 Oe) magnetic relaxation of the FC curve at 150 K is faster (over coming spin pinning) than the slow relaxation (strong domain wall pinning) at 10 K. The in-field magnetization at 150 K switched from positive to negative for $t_w$ > 180 s and wait-in field $H_C$ value (-2150 Oe) becomes smaller than the $H_C$ of ZFC curve (-2533 Oe). This property can be used in magnetic switching/sensor applications. The decrease of $H_C$ at 150 K with the increase of waiting time at -2150 Oe is characteristically opposite with respect to the increment of $H_C$ at 10 K.

We used PC6 to study aging effect on the ZFC-M(H) loop of the composite sample at 10 K with the field sequence +70 kOe to 0 Oe and M($t_w$) measurement for $t_w$ = 3600 s was carried out at 0 Oe (P1). This is followed by resumption of the M(H) measurement down to -7 kOe (P2), where the sample was waited to record the second M($t_w$) curve. Then, recording of M(H) curve was continued down to field at -70 kOe and reversed back to +7 kOe (P3) where third M($t_w$) data were recorded. Finally, M(H) curve continued for fields up to +20 kOe. A similar M(H) measurement protocols were used to record the M($t_w$) curves at 10 K after cooling the sample in the presence of +70 kOe from 300 K. The same experiments were carried out at relatively higher temperature 100 K at field points 0 Oe and ± 4 kOe. Fig. 11 (a-b) shows the recorded ZFC and FC-M(H) curves at 10 K and 100 K. The M($t_w$) curves at points P2 and P3 started to relax towards the magnetization state at zero field. The behavior is consistent to the reverse torque acting on the spins when the field is reduced to zero and rotate over a time toward the positive or negative direction depending on the local easy-axis orientation [20, 37]. The FC loop at 10 K shows nearly symmetric widening along the field and magnetization axes. This showed field cooled induced enhancement of coercivity and magnetization in the composite sample, which showed small exchange bias shift ~ + 64 Oe at 10 K [27]. The field cooled induced widening and exchange bias shift are negligible at 100 K. On the other hand, $M(t_w)$ curves at 10 K and 100 K (Fig. 11(c-d)) followed equation (1) with M(0) values positive and negative for points P2 and P3, respectively. The $M(t_w)$ curves, measured on ZFC and FC-M(H) curves, do not show much differences at 10 K and 100 K for point P1 (waiting at 0 Oe). The normalized $M(t_w)$ curves measured on FC-M(H) curves significantly enhanced for field in-wait ± 7 kOe at 10 K and ± 4 kOe at 100 K in comparison to the measurement on ZFC-M(H) curves. The differences between $M(t_w)$ curves measured on FC-M(H) curves substantially decreased at 100 K, unlike the case on ZFC-M(H) curves at 10 K. The $M(t_w)$ curves in positive field side (+7 kOe at 10 K and + 4 kOe at 100 K) are found to be higher than their counter parts in the negative field side. This indicates the effect of cooling field induced unidirectional anisotropy on interfacial spin ordering [37] and it is reflected in the variation of time constants. Fig. 11(e-f) shows the time constants ($\tau_1$, $\tau_2$) at different fields in-wait in FC process are larger than that in ZFC process both at 10 K and 100 K. This shows a slowing down of spin relaxation due to cooling field induced nucleation of clusters.

Finally, we repeated the measurement of M(H) curves from +70 kOe to -15 kOe at 10 K (Fig. 12(a)) and +70 kOe to -1 kOe at 300 K (Fig. 12(b)) to examine the training effect in the composite sample. The sample was zero field cooled from 300 K (330 K) before repeating the M(H) measurements at 10 K (300 K) for different field sweeping rate without further heating the sample to higher temperature. The M(H) curves at 10 K are practically independent of the field sweeping rate, whereas M(H) curves at 300 K showed magnetic relaxation at higher fields. Most of the systems with training effect have shown decrease of coercivity [11, 38], but coercivity of the composite sample is independent of sweeping rate in the temperature range 10 K to 300 K.

## 4. Conclusions

The ferrimagnetic $Co_{1.75}Fe_{1.25}O_4$ ferrite and its composite with non-magnetic $BaTiO_3$ (BTO) particles are modeled in core-shell spin structure. The existence of intrinsic spin disorder inside the ferrimagnetic ferrite particles is confirmed by exchange bias, memory and relaxation effects. The magnetic memory effect in ferrite particles dominated at higher temperatures, unlike observation of exchange bias effect at low temperature. The magnetic exchange interactions between ferrimagnetic particles are diluted and modified due to presence of intermediating non-magnetic BTO particles. However, basic magnetic properties (blocking of ferrimagnetic clusters, non-equilibrium ferrimagnetic state, memory, exchange bias and aging) of the ferrite particles are retained in the BTO matrix of composite sample. The slow spin dynamics low temperature due to strong spin pinning/inter-cluster interactions becomes faster on increasing the temperature close to the spin freezing/blocking temperature of the samples at $\geq 300$ K (due to increasing fraction of non–interacting/paramagnetic spins). The fraction of paramagnetic spins in composite sample is found to be more (showing pronounced memory effect and faster spin relaxation) than that in the ferrite sample that exhibited relatively small memory dip and slow relaxation. The relaxation of magnetization during field off conditions of the temperature and field dependence of magnetization curves confirmed two relaxation mechanisms in both the samples. The fast relaxation process (initial stage of the relaxation) is attributed to loosely bound shell/interfacial spins, whereas the slow relaxation process (later stage of the relaxation) is attributed to strongly interacting core/interior spins in the systems. The M(H) curves are not much affected by the variation of field sweeping rate, but magnetic state and coercivity of the samples are strongly dependent on in-field or off-field waiting time during M(H) measurements. We showed various options for tuning the ferrimagnetic state and parameters, irrespective of the magnetic ferrite and it's composite in non-magnetic matrix. The tuning capability of ferrimagnetic parameters is promising for technological applications. It may be generalized by applying similar protocols of measurements on different systems.

**Acknowledgment**

RNB acknowledges the Research support from UGC (F. No. F.42-804/2013 (SR)), Govt. of India and UGC-DAE- CSR (No M-252/2017/1022), Gov. of India for carrying out this research work. RNB also acknowledges the PPMS facility at CIF, Pondicherry University.

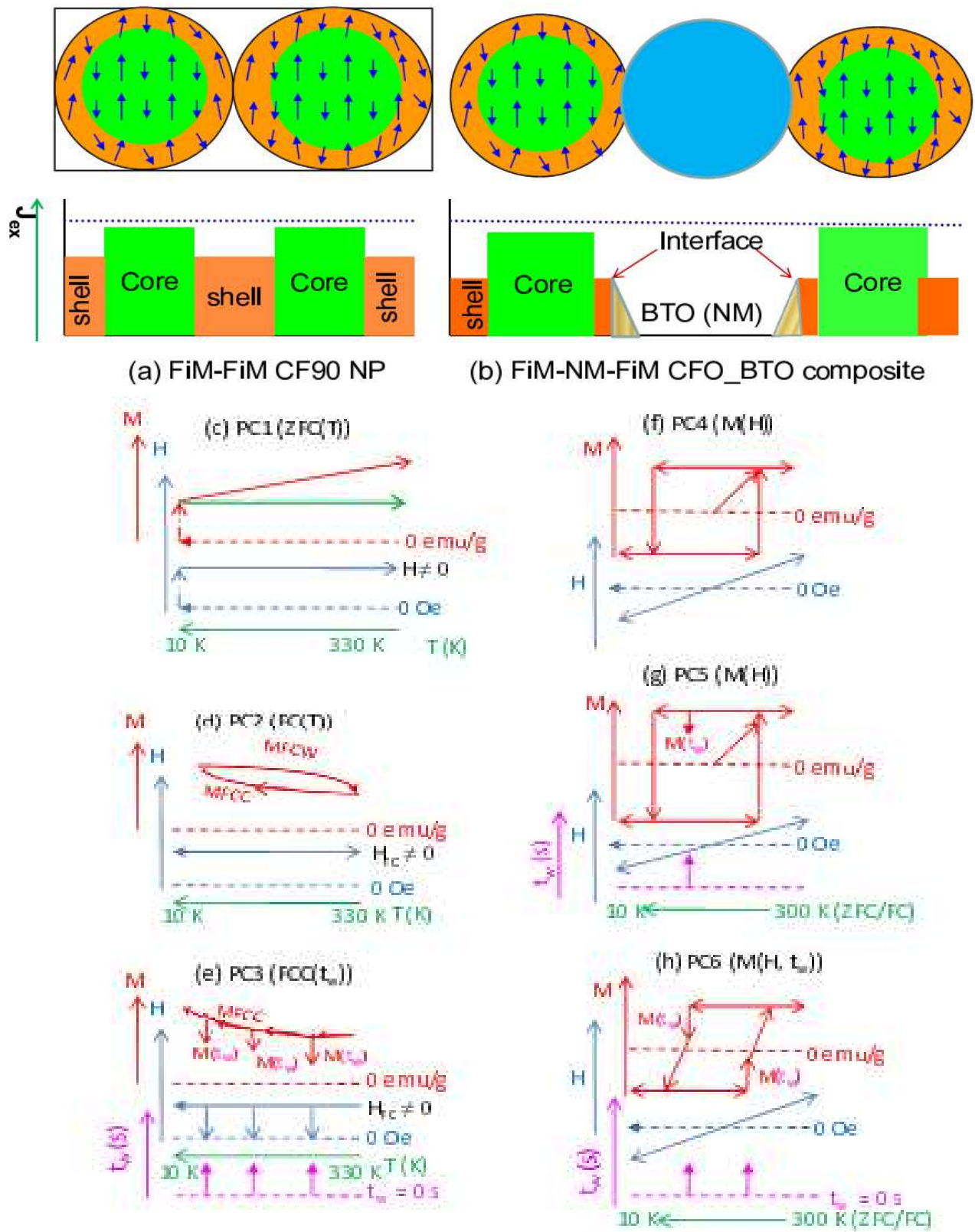

Fig. 1 Schematic of the core-shell spin structure for CF90 (a) and CFO-BTO composite (b) with space variation of exchange interaction ($J_{ex}$). Measurements protocols for temperature (c-e) and field (f-h) dependence of Magnetization, where new protocols were incorporated for recording the time response of the magnetization during field off condition of the magnetization curves. Details described in the text.

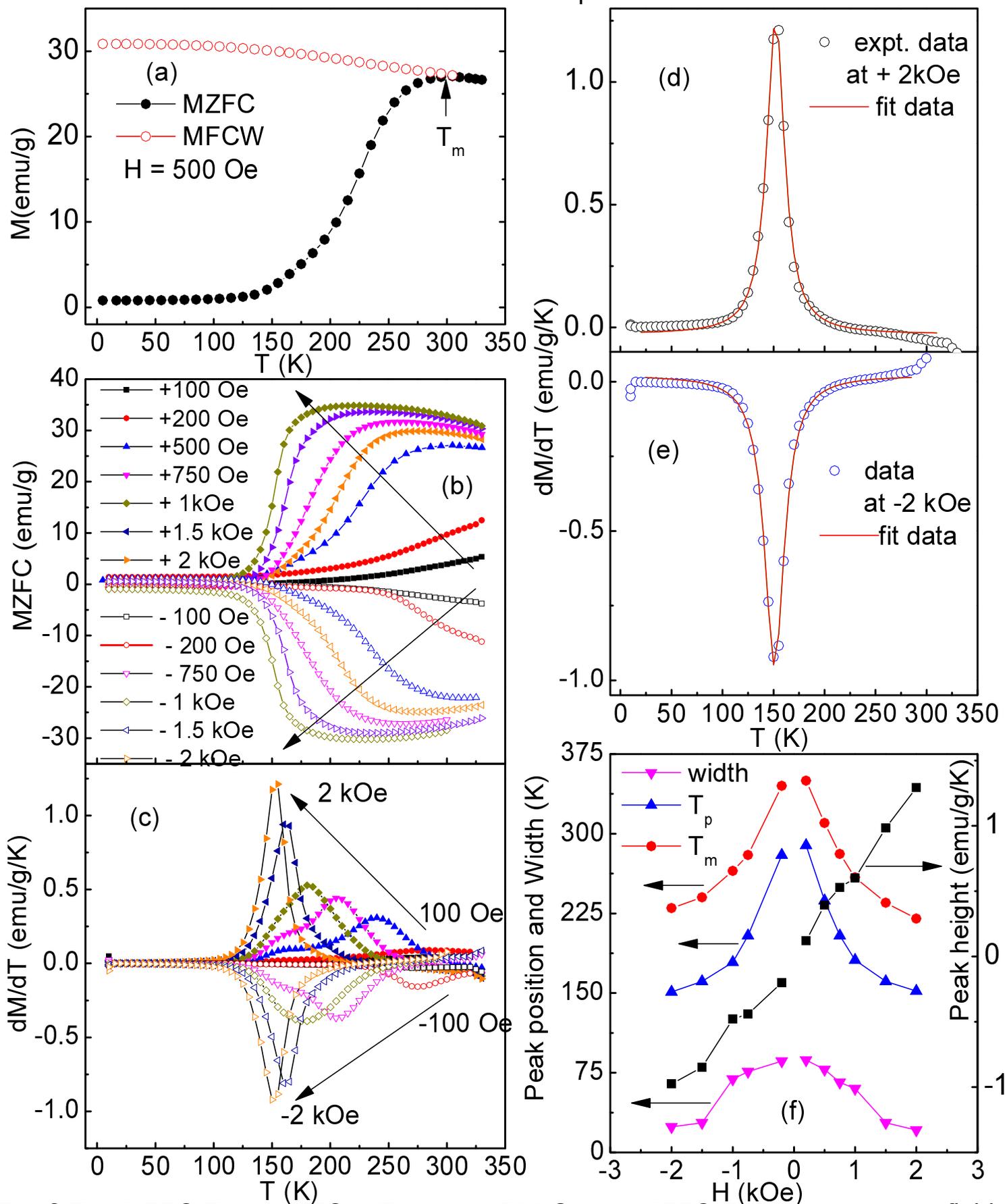

Fig. 2 The MZFC(T) and MFCW(T) data at 500 Oe (a), MZFC(T) data at selected fields (b), first order derivative of the MZFC(T) data in (b), i.e., dM/dT curves (c), fit of dM/dT curves at +2 kOe (d) and at -2 kOe (e). The peak parameters at different fields (f).

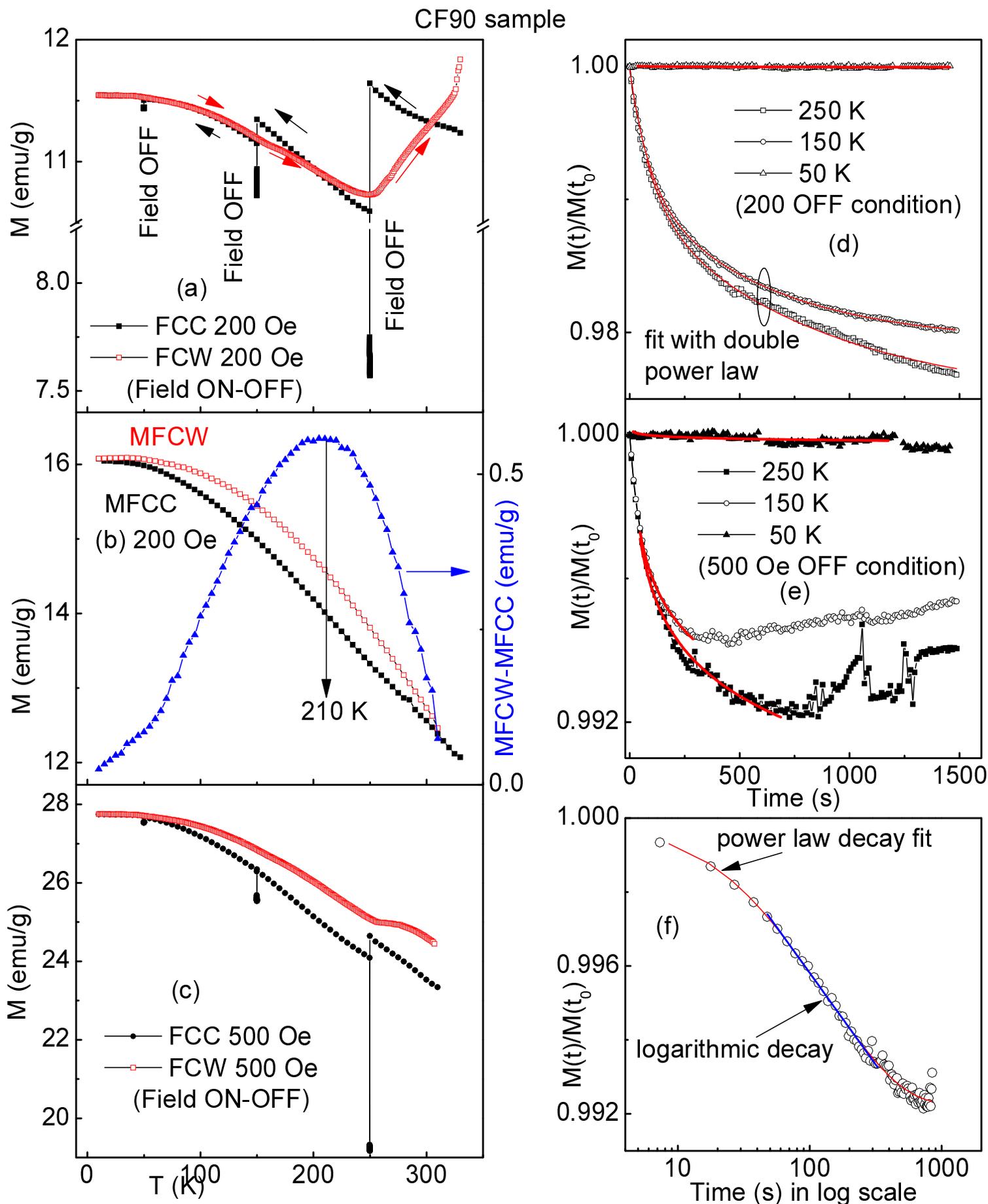

Fig. 3 MFCC(T) with intermediate field OFF sand MFCW(T) at 200 Oe (a). MFCC(T) and MFCW(T) curves at 200 Oe (b) and difference beteween MFCW and MFCC curves at right scale of (b). M(t) data (normalized) during field off at 200 Oe (d) and 500 Oe (e).

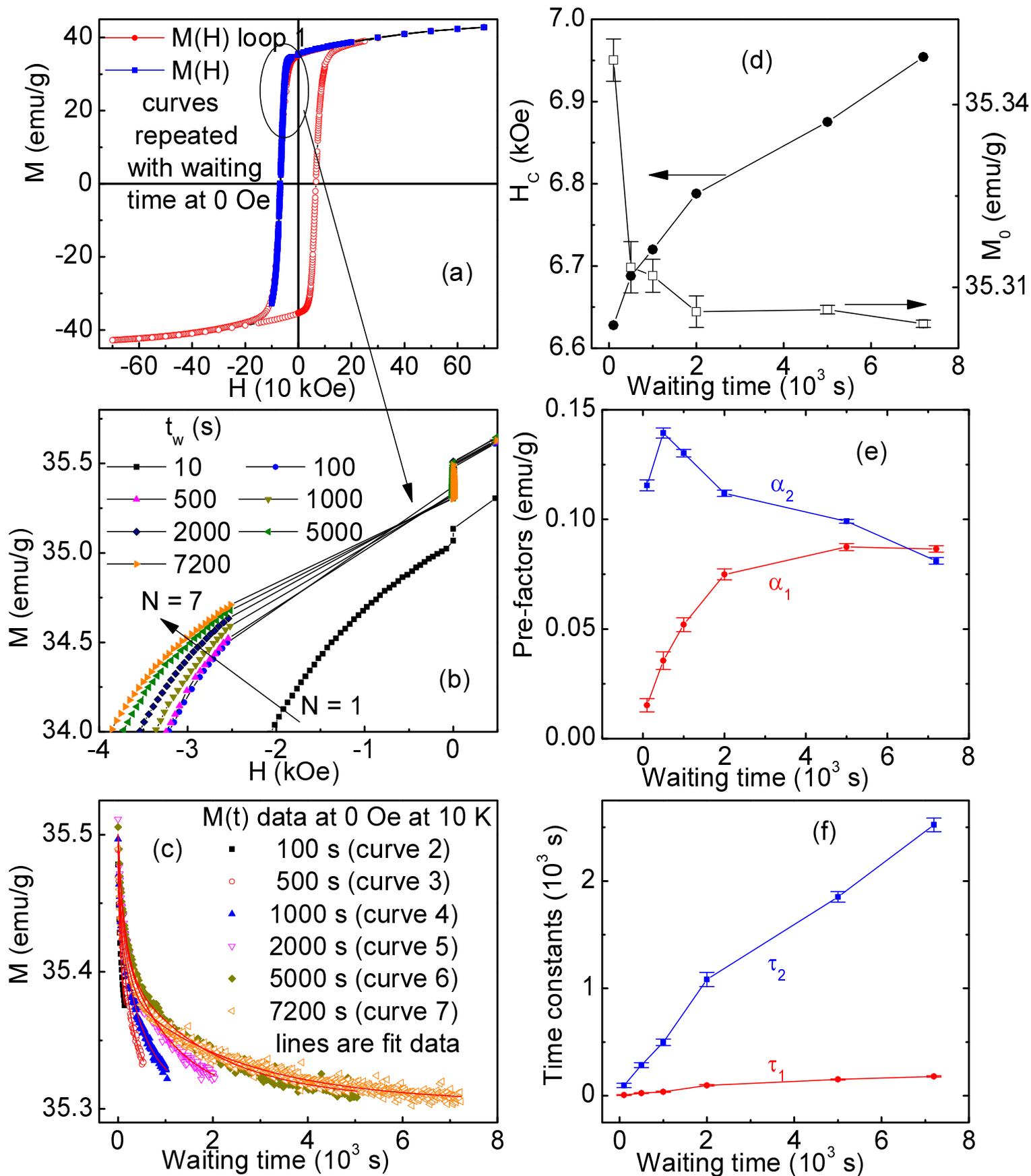

Fig. 4 M(H) loop within +- 70 kOe and repetition of M(H) curves from +70 kOe to -15 kOe with recording of magnetization for different waiting time at 0 Oe (a). Variation of M(H) curves after waiting time (b). M(t) data during waiting time, along with fit data (c). Waiting time dependence of coercivity and initial magnetization (d) and fit parameters of CF90 sample at 10 K (e-f).

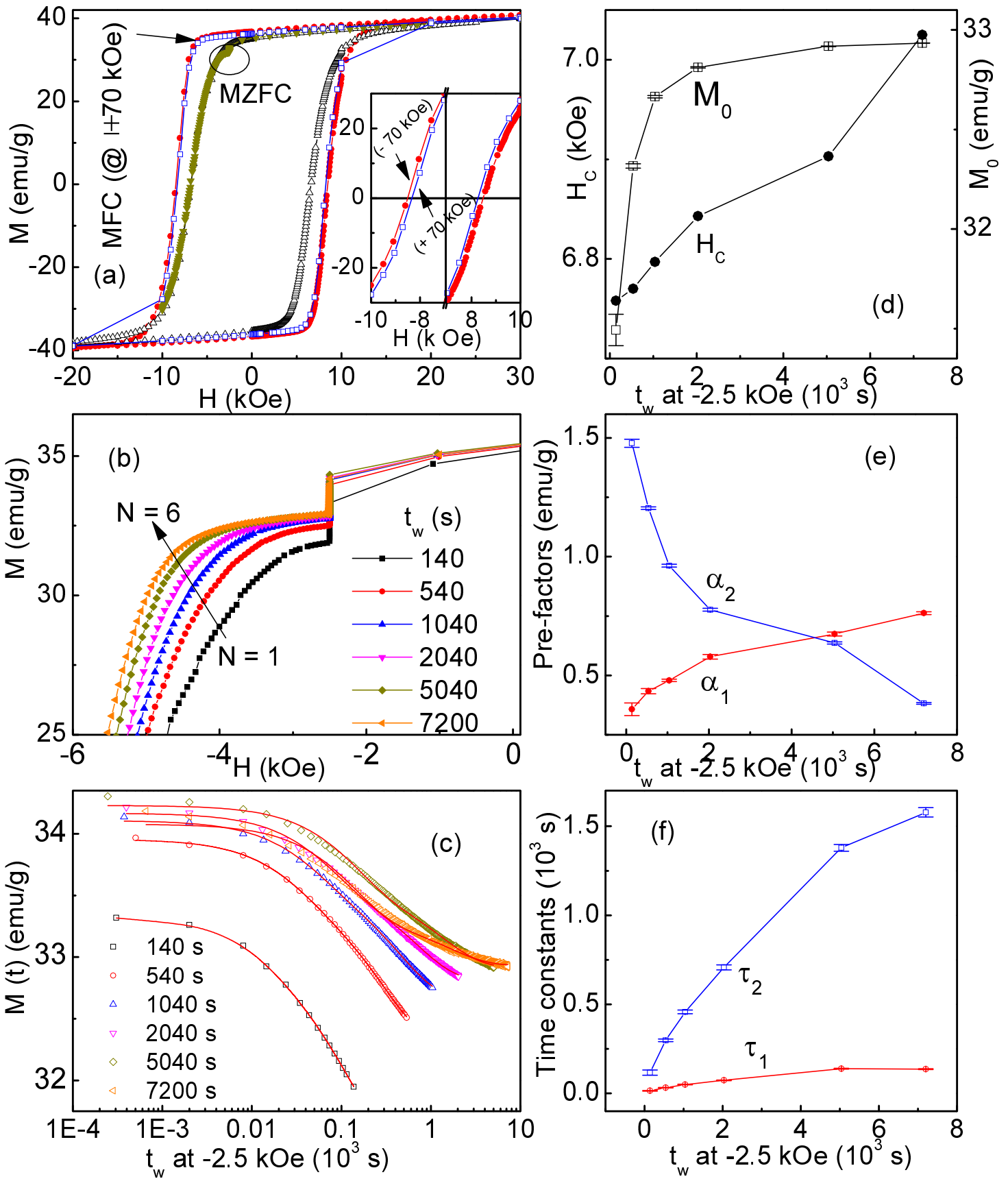

Fig. 5 (a) The M(H) loop in ZFC and FC @±70 kOe modes at 10 K and inset shows the difference in FC loop under + 70 kOe and - 70 kOe. The M(H) curve in ZFC mode was repeated with different waiting time at - 2.5 kOe and the effect is shown in (b). The fit of M(t) data during waiting time (c) and the variation of fit parameters are shown in (d-f).

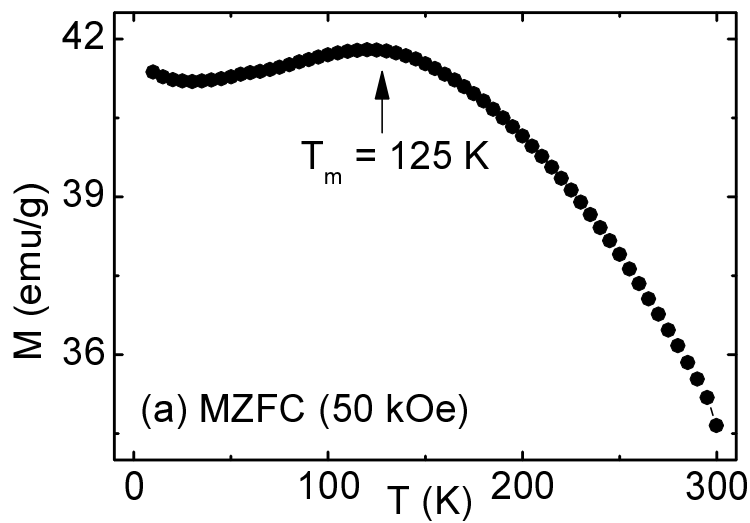
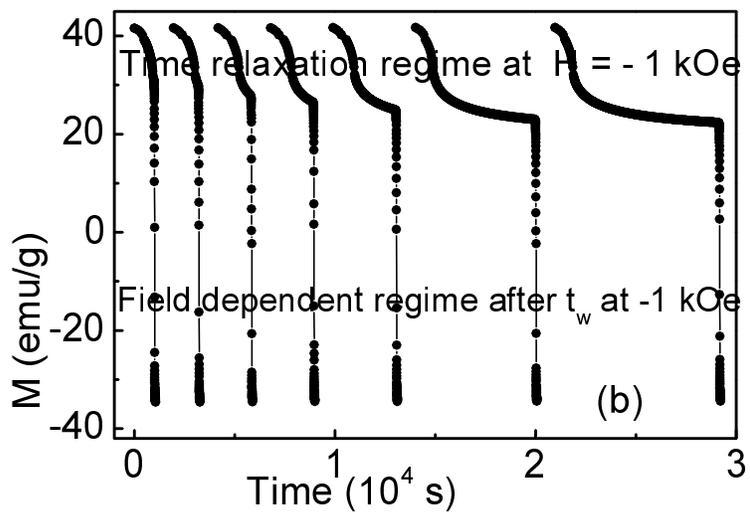
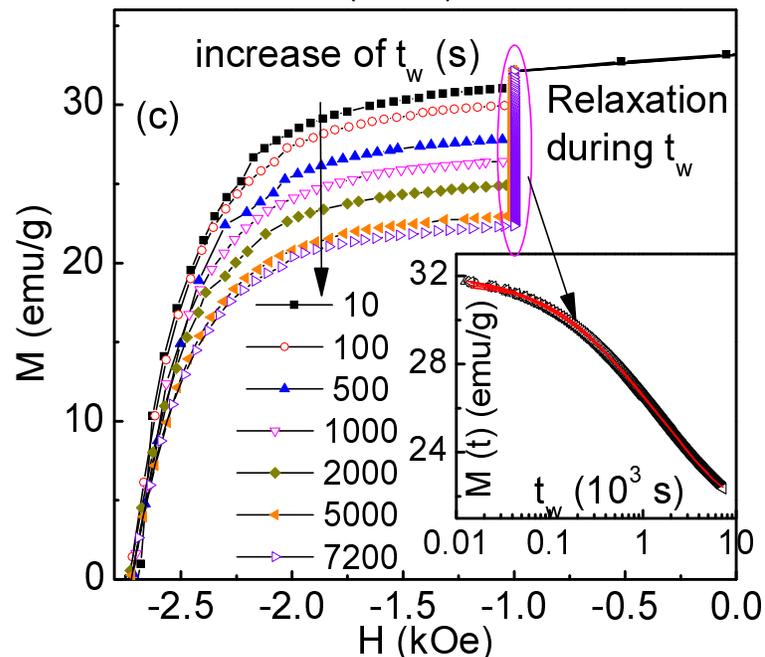
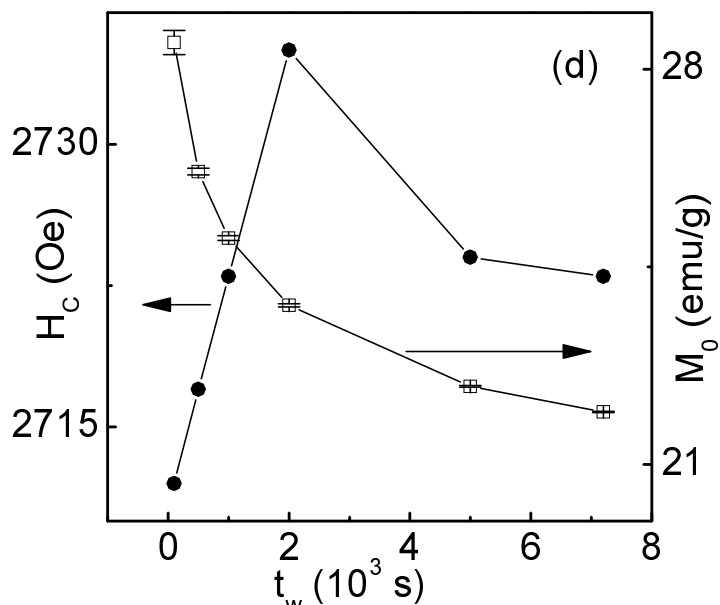
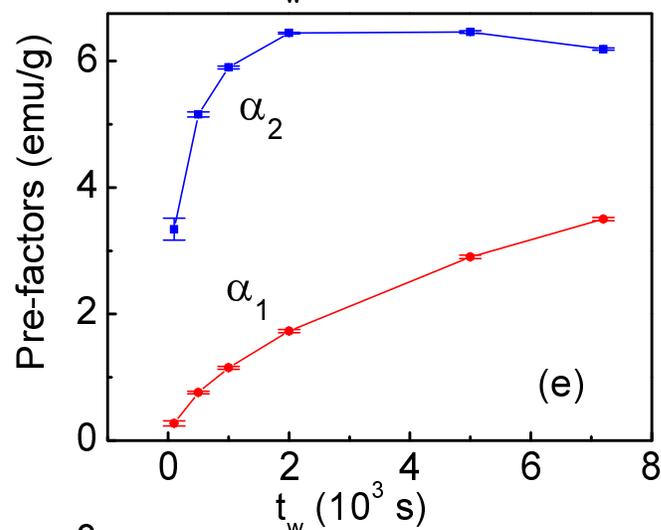
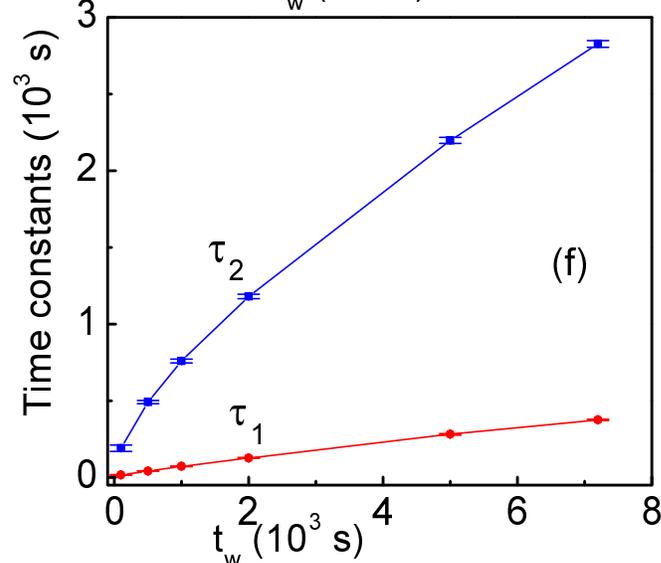

Fig. 6 The MZFC(T) curve measured at 50 kOe (a). The time relaxation and field dependence regime of magnetization in the M(H) curves, which have been repeared with different waiting time at -1 kOe at measurement temperature 150 K(b). The M(H) curve is affected by the $t_w$ at -1 kOe (c). Fit of the M(t) data using exponential law (inset of c) and fit parameters (d-f).

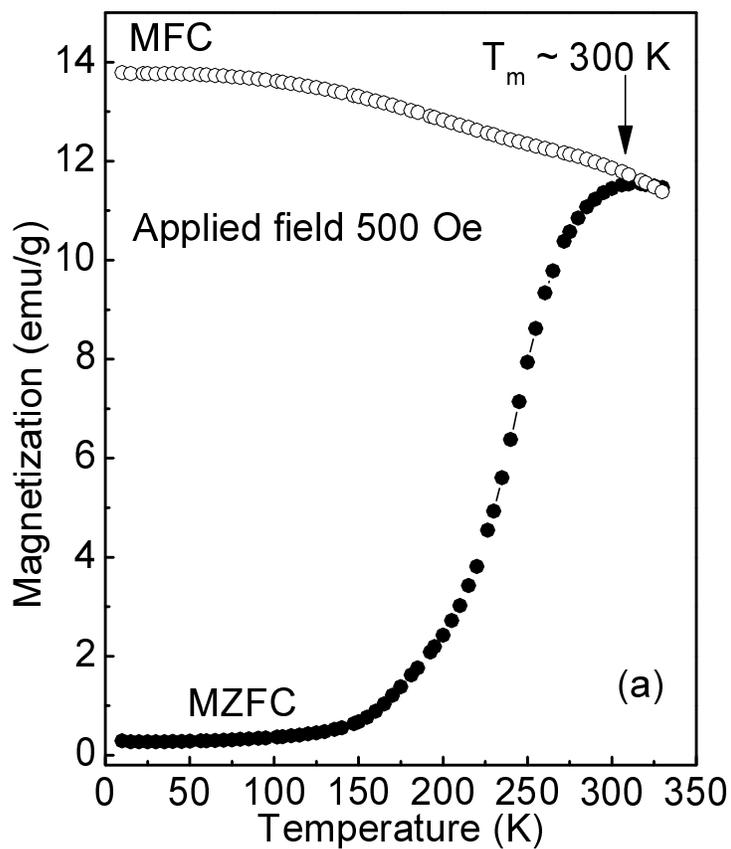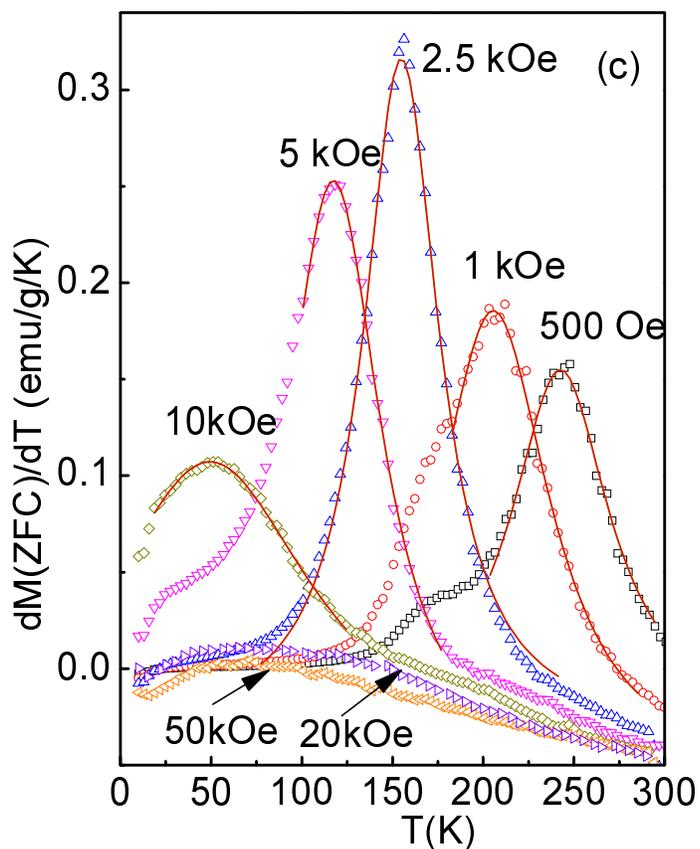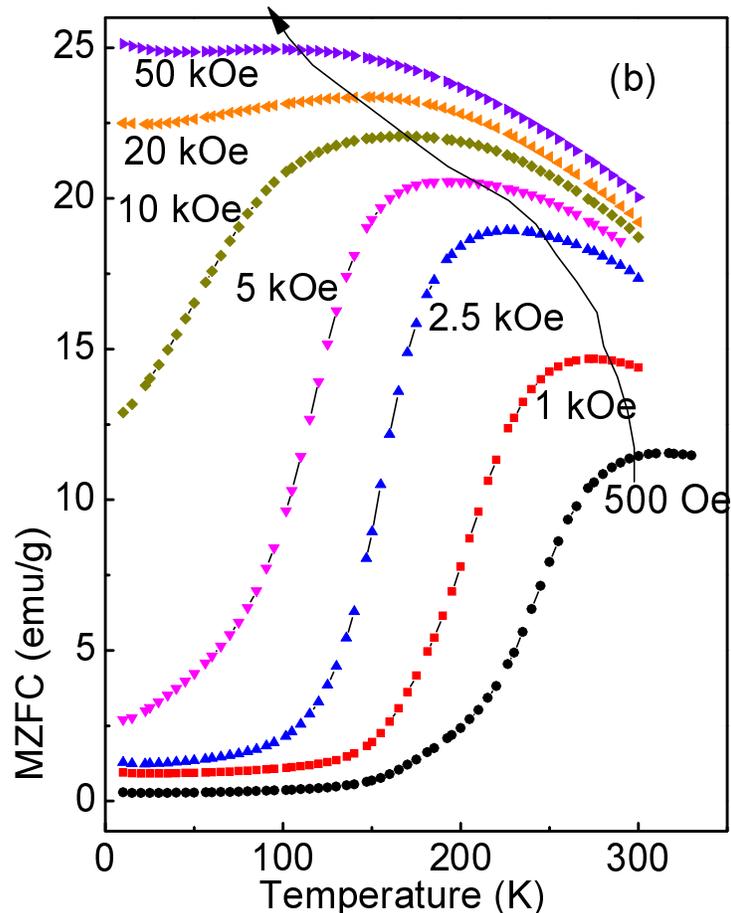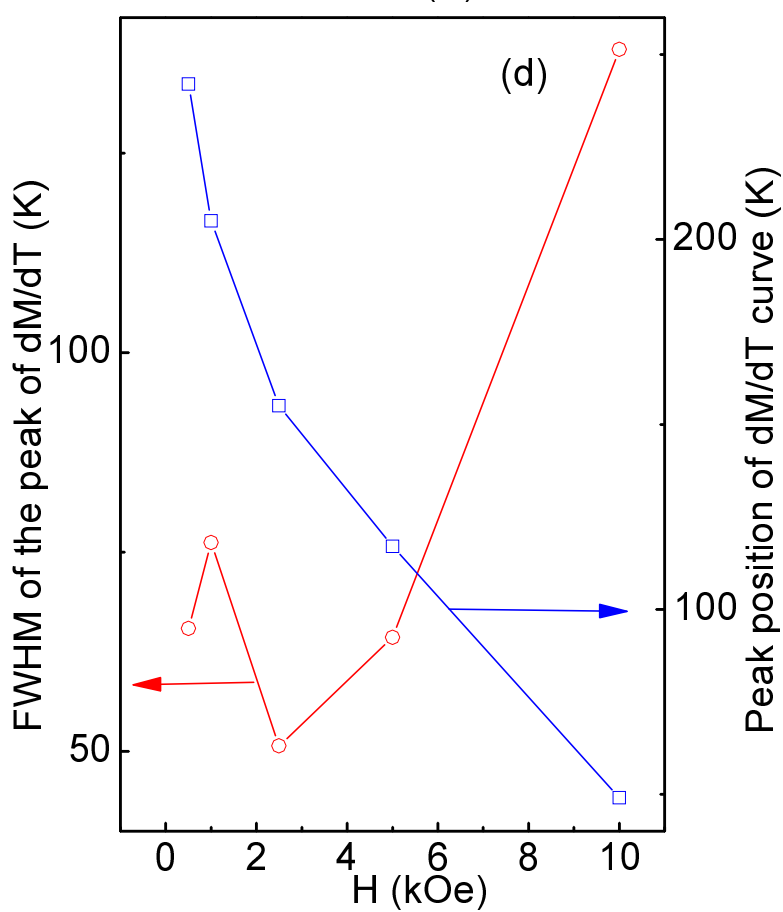

Fig. 7 MZFC(T) and MFC(T) curves at 500 Oe (a). The MZFC (T) curves at different applied fields (b). The temperature derivative of MZFC curves (in b) along with fit of the peaks with Lorentzian shape (c) and variation of the peak parameters with field (d).

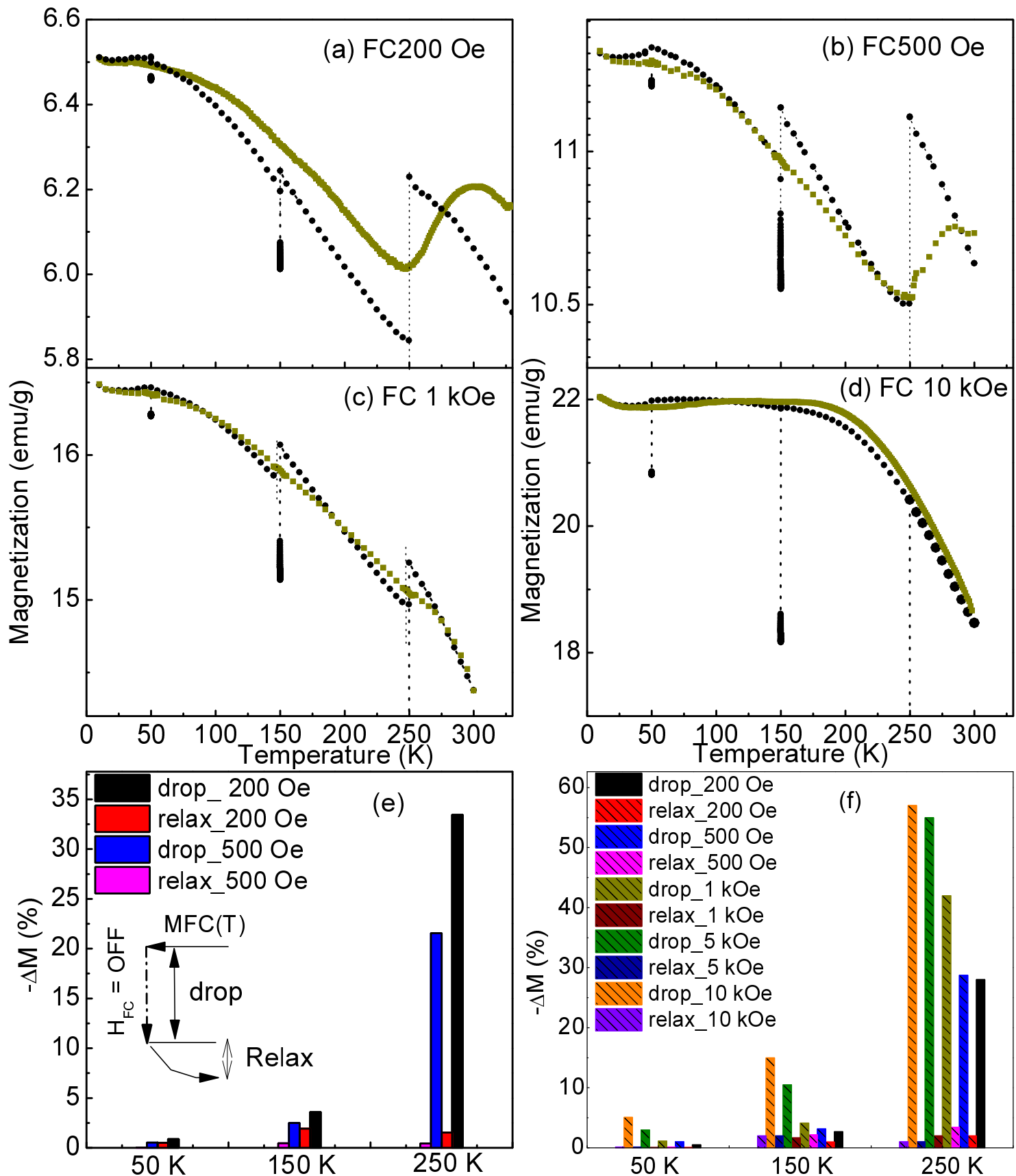

Fig. 8 MFCC(T) and MFCW(T) curves (a-d) of the CF80_BTO composite at selected magnetic fields with field OFF at 250 K, 150 K and 50 K during FCC mode and field continuity in FCW mode (a-d). The drop and relaxation of magnetization in field OFF condition, as sketched in the inset of (e), at 250 K, 150 K and 50 K are compared in bar diagram for CF90 sample (e) and CF80_BTO compoite sample (f).

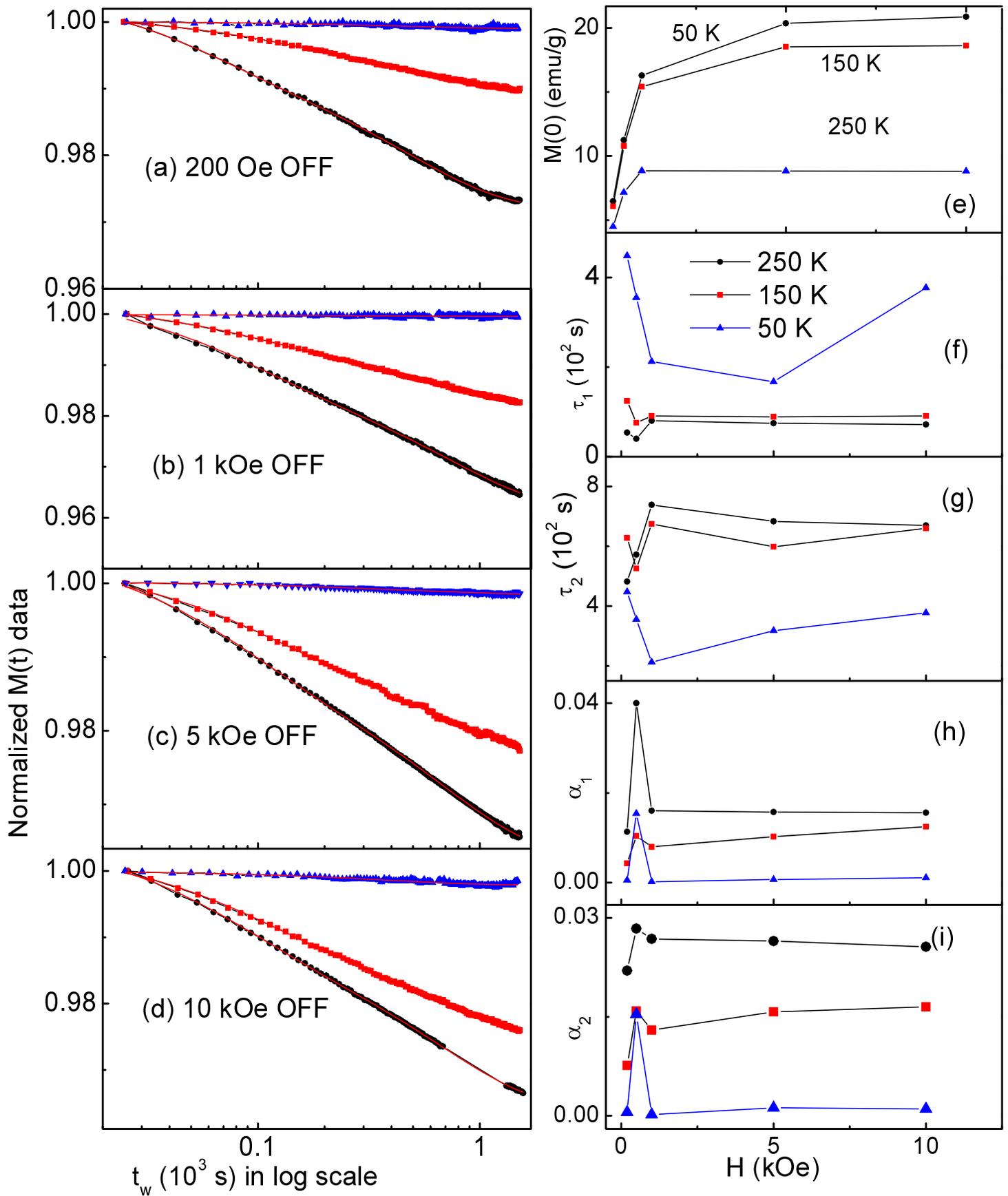

Fig. 9 Fit of the normalized M($t_w$) data of the composite sample(a-d), recorded at field OFF condition during MFCC(T) measurement, using equation (1). The fit parameters are shown in (e-i).

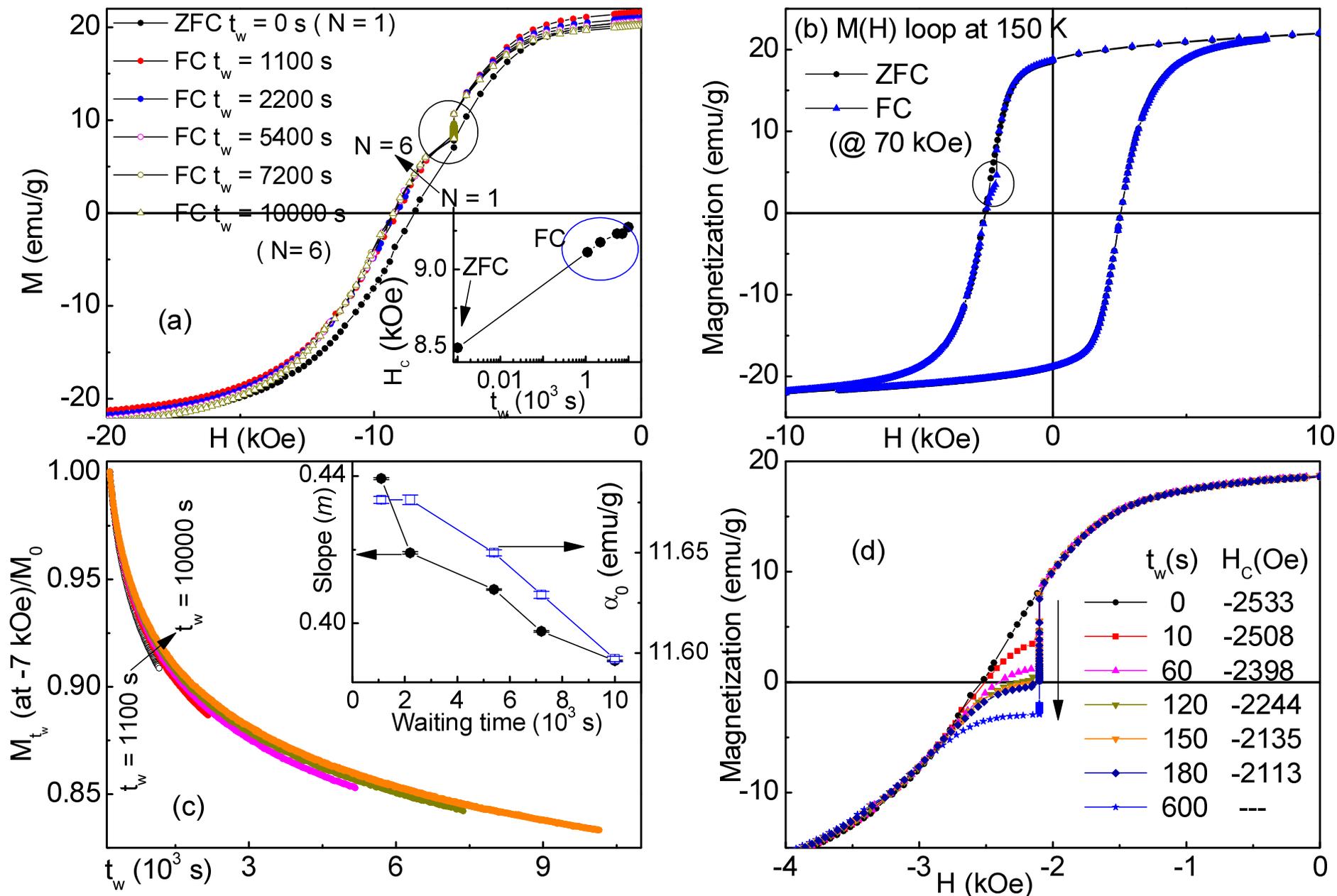

Fig. 10 M-H curves at 10 K under ZFC and FC modes (a). In FC mode, the $M(t_w)$ curves were recorded at -7 kOe, where $t_w$ affected on FC coercivity (inset of a). Similar measurements of M(H) curves under ZFC and FC modes were carried out at 150 K for wait-in field -2100 Oe (b). The in-field $M(t_w)$ curves at 10 K are fitted with logarithmic decay function (c) with fit parameters (inset). The effect of $t_w$ on FC-coercivity and M(H) curves are shown in (d).

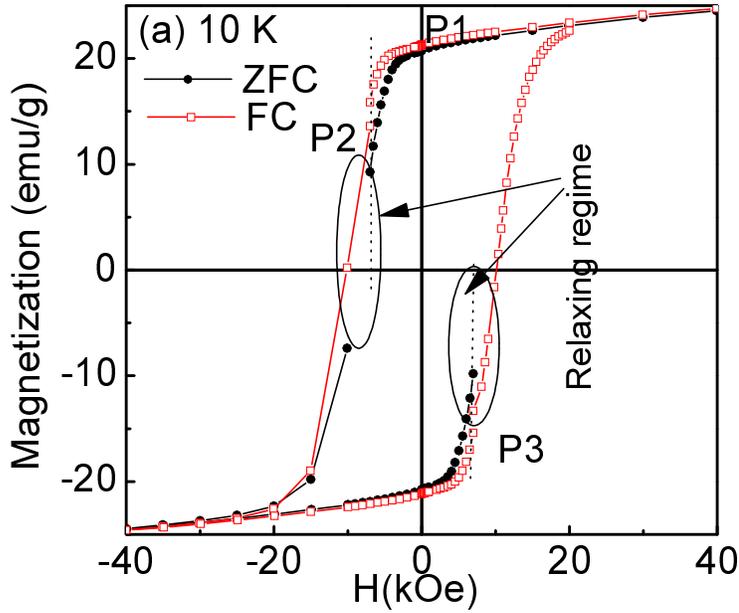
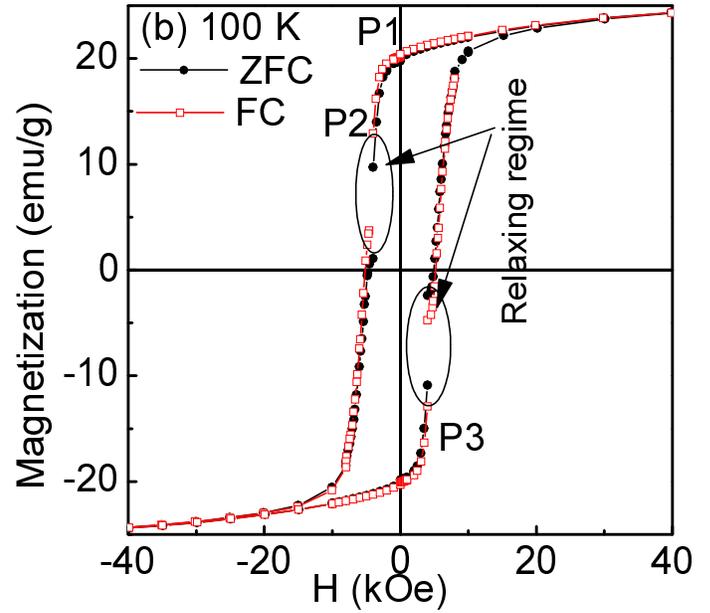
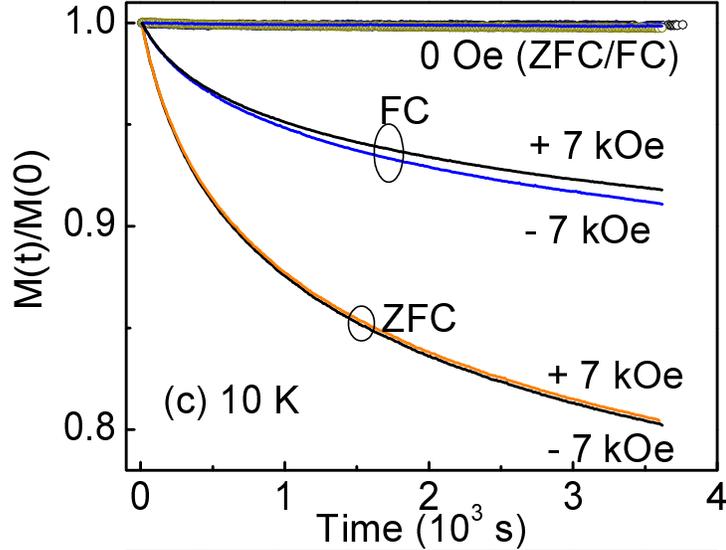
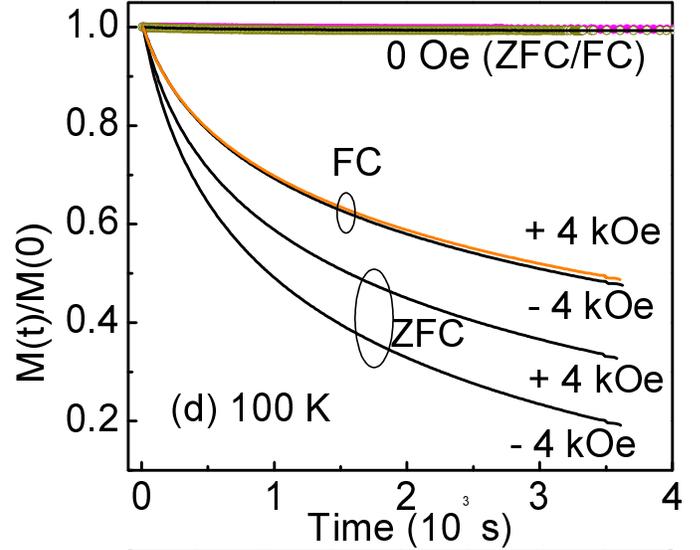
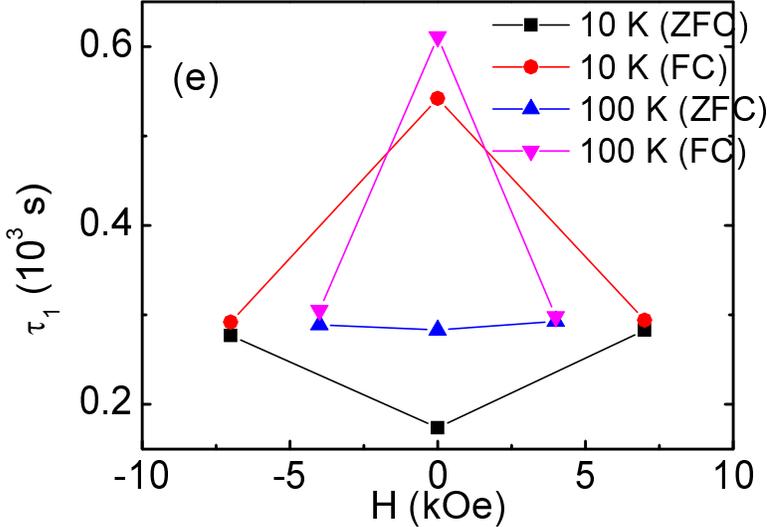
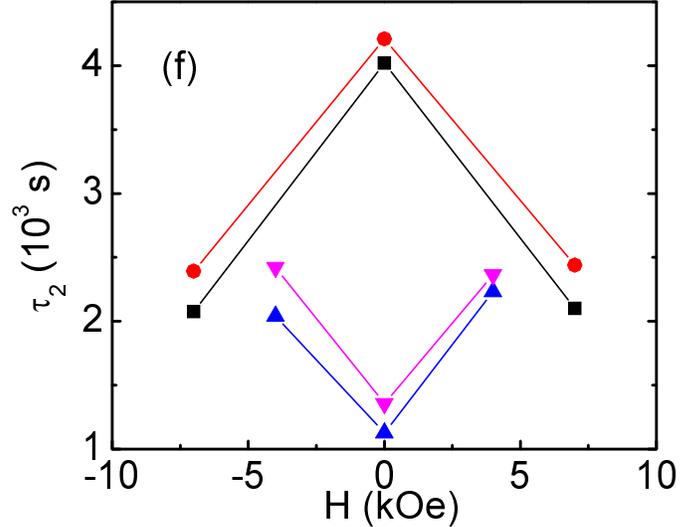

Fig. 11 ZFC and FC loop at 10 K (a) and 100 K (b) with intermediate walting ftime 1 hr for recording M(t) data at points P1, P2 and P3 while field was cycling. The M(t) data (normalized) at 10 K (c) and 100 K(d) were fitted by power law with two time constants ($\tau_1$ and $\tau_2$). The fit values of $\tau_1$ (e) and $\tau_2$ (f) are plotted with constant measurement fields.

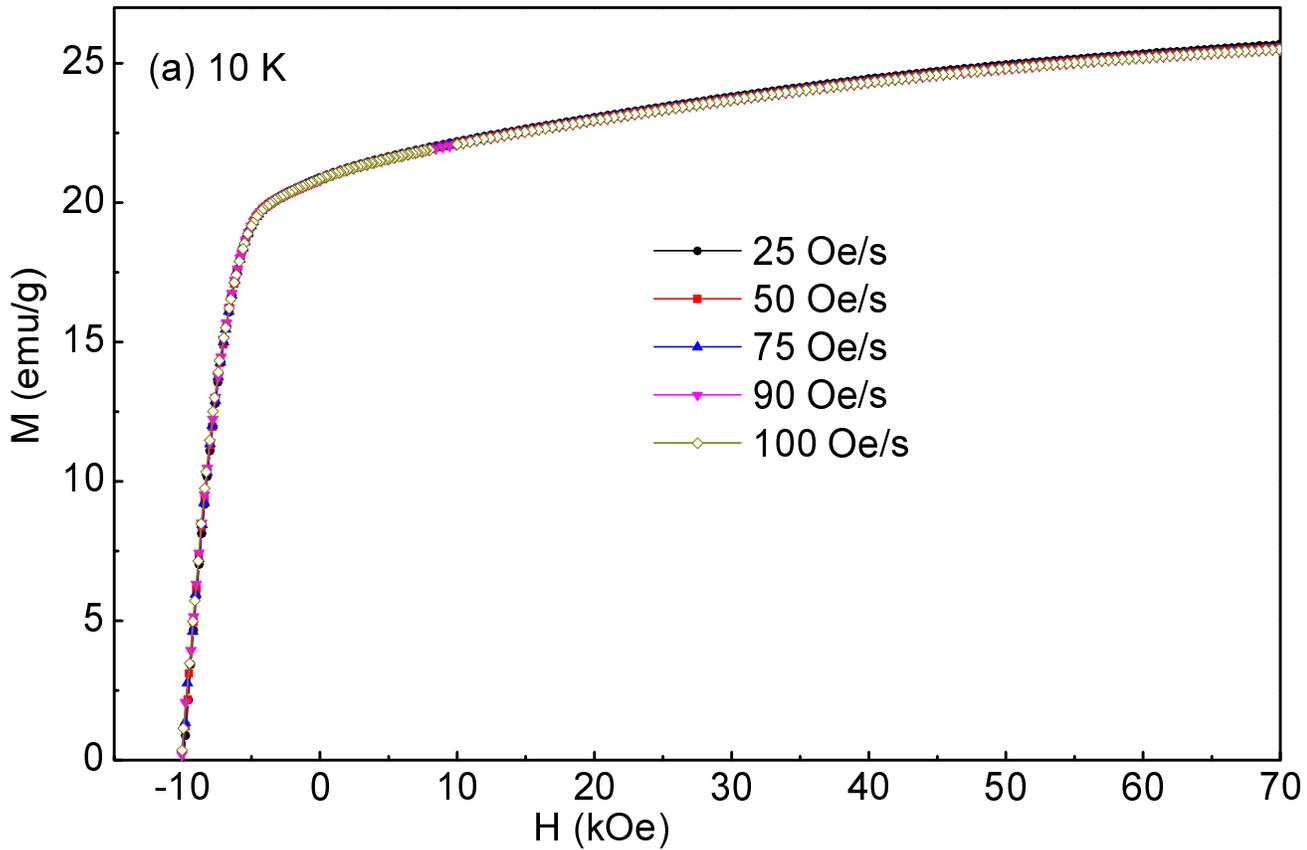
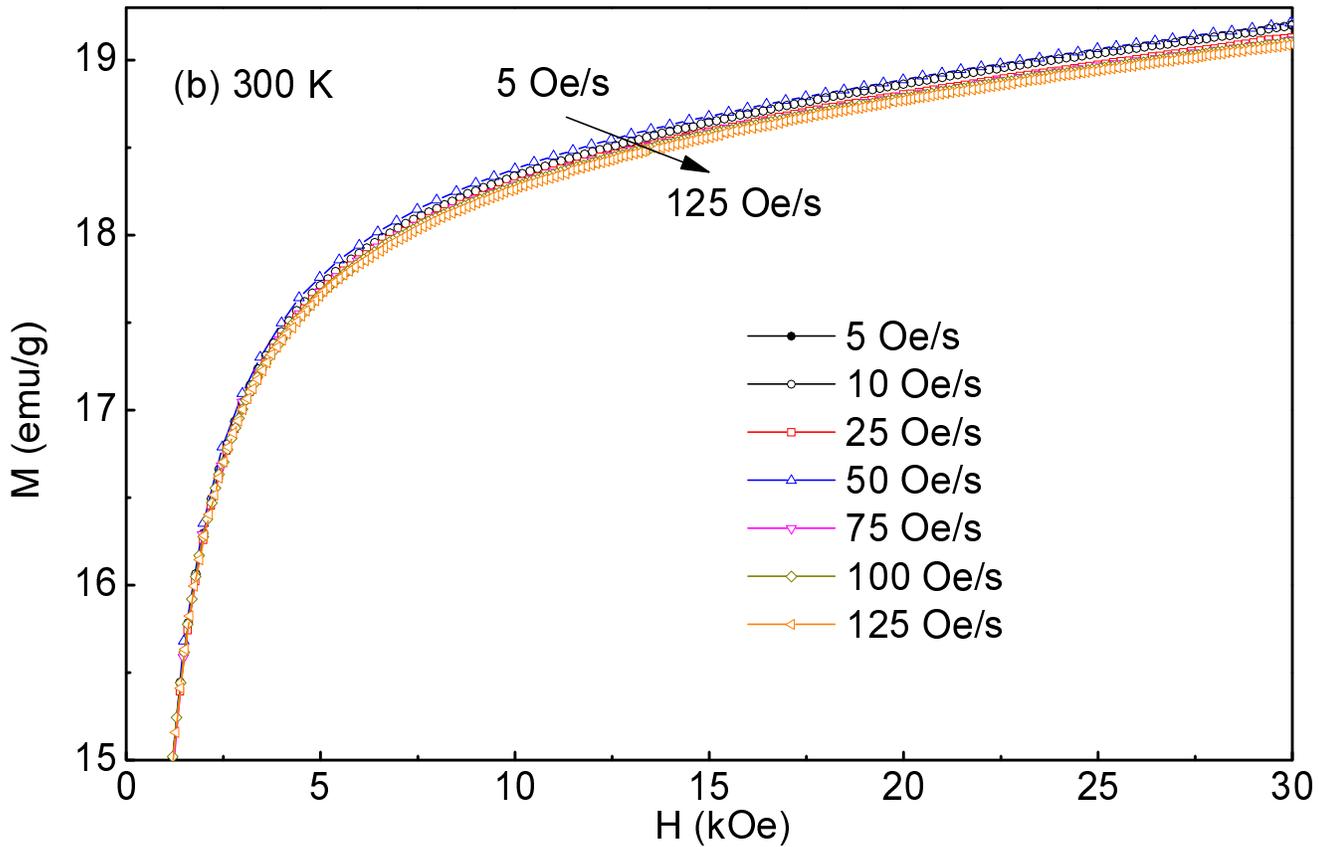

Fig. 12 M(H) curve measurement at 10 K (a) and 300 K (b) was repeated with different field sweeping rate for composite sample.